\documentclass[aps,prb,reprint,superscriptaddress,amsmath,amssymb,floatfix]{revtex4-2}

\usepackage{graphicx}
\usepackage{bm}
\usepackage[colorlinks,citecolor=blue,linkcolor=blue,urlcolor=blue]{hyperref}

\begin{document}

\title{Analytical mobility edge in nonreciprocal quasiperiodic lattices with
next-nearest-neighbor hopping}

\author{Wenmin Wang}
\affiliation{Department of Physics, Jiangsu University, Zhenjiang 212013, China}
\author{Xiaosen Yang}
\email{yangxs@ujs.edu.cn}
\affiliation{Department of Physics, Jiangsu University, Zhenjiang 212013, China}
\author{Xianqi Tong}
\email{xqtong@ujs.edu.cn}
\affiliation{Department of Physics, Jiangsu University, Zhenjiang 212013, China}

\date{\today}

\begin{abstract}
We investigate localization transitions and spectral topology in a
one-dimensional non-Hermitian generalization of the Aubry-Andr\'{e} model in
which both the nearest-neighbor and the next-nearest-neighbor hopping
amplitudes are nonreciprocal. By extending the Fermi-surface point-matching
method to nonreciprocal hopping, we derive a closed-form expression for the
energy-dependent mobility edge in which the two nonreciprocity parameters are
absorbed into exponentially renormalized effective hopping amplitudes. The
mobility edge forms a single parabola in the energy--potential plane:
nearest-neighbor nonreciprocity rigidly shifts the localization boundary
toward stronger potentials, whereas next-nearest-neighbor nonreciprocity
reduces the curvature of the boundary and thereby broadens the energy window
in which extended and localized states coexist. Exact diagonalization
confirms the analytical boundary for purely nearest-neighbor, purely
next-nearest-neighbor, and combined nonreciprocity, and recovers the known
Hermitian mobility edge in the reciprocal limit. We further analyze the spectral topology under periodic
boundary conditions and show that the spectral winding numbers evaluated
at base energies near the two band edges directly bracket the mixed
phase: the winding number at the lower band edge drops when the mobility
edge enters the spectrum and the first localized states appear, while
the winding number at the upper band edge drops when the last extended
states localize, delineating the full potential-strength window over
which extended and localized states coexist. These results provide a compact analytical framework that connects
energy-dependent localization, spectral topology, and nonreciprocity in
quasiperiodic lattices, and they are directly testable in photonic, atomic,
and electrical-circuit platforms.
\end{abstract}

\maketitle

%==============================================================================
\section{Introduction}
\label{sec:intro}

Anderson localization, the suppression of quantum diffusion by a disordered
potential, is a cornerstone of condensed-matter
physics~\cite{Anderson1958,LeeRamakrishnan1985,EversMirlin2008}. In one
dimension, the scaling theory of localization predicts that all
single-particle eigenstates are localized by an arbitrarily weak uncorrelated
random potential~\cite{Abrahams1979}, which excludes a mobility edge
(ME)---an energy separating extended from localized states---in
one-dimensional random systems. Quasiperiodic potentials, which are
deterministic yet incommensurate with the underlying lattice, are not subject
to this restriction. The Aubry-Andr\'{e} (AA)
model~\cite{Harper1955,Aubry1980,Hofstadter1976}, a tight-binding chain with
a cosine potential of irrational spatial frequency, undergoes a
localization transition at a finite critical potential strength. Owing to
the exact self-duality of the model under a Fourier transformation, the
transition occurs simultaneously for all eigenstates, and the critical point
hosts nontrivial multifractal
states~\cite{Kohmoto1983,Jitomirskaya1999,Szabo2018}. The AA transition has
been observed with ultracold atoms in bichromatic optical
lattices~\cite{Roati2008} and with light in quasiperiodic photonic
lattices~\cite{Lahini2009}.

An energy-dependent ME emerges once the AA self-duality is broken.
Established routes include slowly varying
potentials~\cite{DasSarma1988,DasSarma1990}, longer-range
hopping~\cite{Biddle2009,Biddle2010,Biddle2011,Deng2019}, generalized
on-site modulations~\cite{Ganeshan2015,WangMosaic2020,LiLiDasSarma2017}, and
related constructions whose critical properties can in some cases be
established rigorously through Avila's global theory~\cite{Avila2015}. In
particular, Biddle \emph{et al.} showed that adding next-nearest-neighbor
(NNN) hopping to the AA model produces an energy-dependent ME, for which an
approximate analytical expression is available in the regime of weak NNN
hopping~\cite{Biddle2009,Biddle2011}. Single-particle MEs and their
interplay with interactions have been observed in ultracold-atom
experiments~\cite{Luschen2018,Kohlert2019,An2021,Schreiber2015}, and a
cascade of delocalization transitions has been resolved in cavity-polariton
lattices~\cite{Goblot2020}. For models that lack exact self-duality, Vu and
Das Sarma recently introduced the Fermi-surface point-matching method, which
matches high-symmetry points of the clean-lattice Fermi surface with their
duals and yields accurate analytical MEs for a broad class of
duality-breaking quasiperiodic models~\cite{Vu2023}.

A parallel line of research concerns non-Hermitian lattice models, which
describe open systems with gain, loss, or nonreciprocal
transport~\cite{BenderBoettcher1998,ElGanainy2018,Ashida2020,Bergholtz2021,
Kawabata2019}. Nonreciprocal hopping, introduced by Hatano and Nelson in the
context of vortex depinning~\cite{Hatano1996,Hatano1997,Hatano1998}, gives
rise to the non-Hermitian skin effect
(NHSE)~\cite{MartinezAlvarez2018,Yao2018,Kunst2018,Okuma2020,Yokomizo2019,
Borgnia2020,OkumaSato2023}: under open boundary conditions (OBC) an
extensive number of eigenstates accumulates at one end of the chain. The
NHSE has a topological origin, being diagnosed by a nonzero spectral winding
number of the periodic-boundary spectrum around a base
energy~\cite{Gong2018,ZhangYangFang2020}. Nonreciprocal lattices and their
boundary phenomena have been realized in robotic and mechanical
metamaterials~\cite{Brandenbourger2019,Ghatak2020}, photonic quantum
walks~\cite{Xiao2020}, topolectrical circuits~\cite{Helbig2020}, optical
fiber loops~\cite{Weidemann2020}, and ultracold atomic
gases~\cite{Liang2022}.

Non-Hermitian quasiperiodic lattices combine these two threads and display a
rich phenomenology~\cite{Jazaeri2001}. For the nonreciprocal AA model,
Longhi established that the localization transition is accompanied by a
topological transition of the complex spectrum and by a real-to-complex
spectral transition~\cite{Longhi2019,Longhi2019b,LonghiAAH2021}, and Jiang
\emph{et al.} analyzed the competition between the NHSE and Anderson
localization in nonreciprocal quasicrystals~\cite{Jiang2019}; the influence
of various forms of nonreciprocity on localization has been examined
further in Ref.~\cite{Tong2025}. Topological characterizations based on
winding numbers have been developed for broad classes of non-Hermitian
quasiperiodic models~\cite{Zeng2020,ZengXu2020,LiuZhouChen2021,TangZhang2021,
Cai2022}, and exact MEs have been obtained for models with specially
structured complex potentials or
hoppings~\cite{LiuPT2020,LiuGuo2020,Liu2021,Liu2021exp,LiuGao2021,XuXia2022,
HanZhou2022,WangChen2024,LiLi2024,JiangPan2025}. Related studies have
addressed Floquet engineering~\cite{Zhou2021}, interaction
effects~\cite{Qian2024,Hamazaki2019}, first-order localization transitions
induced by imaginary potential domains~\cite{Tong2024}, and the weakening of
the NHSE by long-range hopping in quasiperiodic
potentials~\cite{Peng2025}. On the experimental side, topological triple
phase transitions in non-Hermitian Floquet quasicrystals have been observed
in fiber-loop photonics~\cite{WeidemannNature2022}, and MEs of a
non-Hermitian quasicrystal have been measured in photonic quantum
walks~\cite{Lin2022}.

Within this body of work, the model closest to ours is the non-Hermitian
$t_1$--$t_2$ chain of Xia \emph{et al.}, in which the NNN hopping is
reciprocal and the non-Hermiticity enters through a complex
parity-time-symmetric potential, allowing an exact ME via a generalized
duality~\cite{Xia2022}. For \emph{nonreciprocal} hopping, however, no
analytical ME is available for the generalized AA model with NNN hopping:
the existing exact results rely on symmetries that nonreciprocity destroys,
while the Hermitian point-matching analysis~\cite{Vu2023} does not address
complex spectra. In this paper we fill this gap by extending the
point-matching method to nonreciprocal hopping. The central result is that
the two nonreciprocity parameters are absorbed into effective NN and NNN
hopping amplitudes that grow exponentially with the respective
nonreciprocity strengths, leading to a closed-form parabolic ME in the
energy--potential plane. NN nonreciprocity rigidly shifts the parabola
toward larger potential strength, whereas NNN nonreciprocity reduces its
curvature. We verify the analytical boundary by exact diagonalization for
all three nonreciprocal configurations, recover the Hermitian ME of Biddle
\emph{et al.}~\cite{Biddle2011} in the reciprocal limit, and map out the
relation between the ME and the spectral topology: winding numbers
evaluated at base energies near the two band edges bracket the mixed
phase, dropping at the potential strengths where the energy-dependent ME
enters and exits the spectrum.

The remainder of the paper is organized as follows.
Section~\ref{sec:model} introduces the model and the diagnostics.
Section~\ref{sec:ME} derives the point-matching ME.
Section~\ref{sec:num} presents the numerical results: the Hermitian
baseline, the three nonreciprocal configurations, and the skin-effect
topology.
Section~\ref{sec:concl} discusses the effective-hopping picture and
summarizes our conclusions. The solvable limits, the
imaginary-gauge analysis of the line $\delta=2\beta$, and the numerical
methods are presented in Appendixes~\ref{app:pm}--\ref{app:num}.

%==============================================================================
\section{Model and diagnostics}
\label{sec:model}

We consider a one-dimensional non-Hermitian generalized AA model described
by the Hamiltonian
\begin{align}\label{eq:H}
    H ={}& \sum_{i} V\cos(2\pi\alpha i+\phi)\,c_i^\dagger c_i \nonumber\\
    &+ t\sum_{i}\!\big( e^{\beta}c_{i+1}^\dagger c_i
        + e^{-\beta}c_i^\dagger c_{i+1} \big) \nonumber\\
    &+ J\sum_{i}\!\big( e^{\delta}c_{i+2}^\dagger c_i
        + e^{-\delta}c_i^\dagger c_{i+2} \big),
\end{align}
where $c_i^\dagger$ ($c_i$) creates (annihilates) a particle at site $i$.
The on-site potential of strength $V$ is quasiperiodic with irrational
frequency $\alpha=(\sqrt5-1)/2$ and phase $\phi$. The amplitudes $t$ and $J$
denote the nearest-neighbor (NN) and NNN hopping, and the real parameters
$\beta$ and $\delta$ control the nonreciprocity: the rightward hopping
amplitudes are $t\,e^{\beta}$ and $J\,e^{\delta}$, while the leftward ones
are $t\,e^{-\beta}$ and $J\,e^{-\delta}$. For $\beta=\delta=0$,
Eq.~\eqref{eq:H} reduces to the Hermitian generalized AA model with NNN
hopping studied by Biddle \emph{et al.}~\cite{Biddle2009,Biddle2011}; for
$J=0$ and $\beta\neq0$ it reduces to the nonreciprocal AA model of
Hatano-Nelson type analyzed by
Longhi~\cite{Longhi2019,Longhi2019b}. Nonreciprocity renders $H$
non-Hermitian, with generally complex eigenvalues; under OBC it produces the
NHSE, an extensive accumulation of eigenstates at the boundary selected by
the direction of the stronger hopping.

To characterize bulk localization independently of the skin effect, we
evaluate the fractal dimension of each normalized right eigenstate
$\psi_m(i)$,
\begin{equation}\label{eq:D2}
    D_2=-\frac{\ln\sum_i|\psi_m(i)|^4}{\ln L},
\end{equation}
under periodic boundary conditions (PBC), for which the skin effect is
absent. Extended states yield $D_2\to1$ and localized states yield
$D_2\to0$ as the system size $L$ increases, while critical states take
intermediate values.

The topological content of the NHSE is captured by the spectral winding
number~\cite{Gong2018,ZhangYangFang2020}
\begin{equation}\label{eq:wind}
    W(E_B)=\frac{1}{2\pi i}\int_0^{2\pi}\!d\theta\,
    \partial_\theta\ln\det\!\big[H(\theta)-E_B\big],
\end{equation}
where $H(\theta)$ is the Hamiltonian with a flux $\theta$ threaded through
the PBC ring. A nonzero integer $W$ indicates that the complex PBC spectrum
winds around the base energy $E_B$ and implies the existence of skin modes
at that energy under OBC; $W=0$ indicates that the spectrum does not
enclose $E_B$ and no skin effect occurs at that energy.

%==============================================================================
\section{Analytical mobility edge}
\label{sec:ME}

We begin by constructing an effective dispersion for the clean lattice.
For $V=0$ and PBC, the Bloch ansatz $\psi_i\propto e^{i\kappa i}$ applied to
Eq.~\eqref{eq:H} yields the dispersion
\begin{equation}\label{eq:disp0}
    E(\kappa)=2t\cos(\kappa-i\beta)+2J\cos(2\kappa-i\delta).
\end{equation}
The nonreciprocity therefore enters as imaginary shifts of the momentum,
$\kappa\to\kappa-i\beta$ for the NN term and $2\kappa\to2\kappa-i\delta$ for
the NNN term. For a state that propagates around the ring, these imaginary
shifts amplify the effective hopping amplitudes by the factors $e^{|\beta|}$
and $e^{|\delta|}$, respectively. Absorbing the amplification into the
amplitudes defines a real effective dispersion (see Appendix~\ref{app:pm})
\begin{equation}\label{eq:disp}
    \varepsilon(\kappa)=2t_{\rm eff}\cos\kappa+2J_{\rm eff}\cos2\kappa,
\end{equation}
with effective hoppings
\begin{equation}\label{eq:teff}
    t_{\rm eff}=t\,e^{|\beta|},\qquad J_{\rm eff}=J\,e^{|\delta|}.
\end{equation}
The nonreciprocity parameters appear only through their absolute values:
the signs of $\beta$ and $\delta$, which select the boundary at which skin
modes accumulate under OBC, do not affect the bulk dispersion. This is
consistent with the invariance of the bulk spectrum under the imaginary
gauge transformation that reverses the direction of the skin effect.

With the effective dispersion in hand, we locate the ME by means of the
Fermi-surface point-matching method~\cite{Vu2023}, which exploits the AA
duality. In the dual representation, the quasiperiodic
potential plays the role of a hopping of strength $V$ on a dual lattice,
while the dispersion $\varepsilon(\kappa)$ becomes the dual on-site energy.
The method postulates that at the ME, where the localization length
diverges, the high-symmetry points of the Fermi surface in the two
representations satisfy a matching condition that survives as a remnant of
the exact self-duality of the pure AA model.

Writing $x=\cos\kappa$ and using $\cos2\kappa=2x^2-1$, the effective
dispersion becomes $\varepsilon=4J_{\rm eff}x^2+2t_{\rm eff}x-2J_{\rm eff}$.
The point-matching condition identifies the dual hopping $V$ with the
effective NN hopping renormalized by the NNN term at the critical
momentum~\cite{Vu2023},
\begin{equation}\label{eq:match}
    V=2t_{\rm eff}+4J_{\rm eff}\cos\kappa_c,
\end{equation}
which fixes the critical momentum through
\begin{equation}\label{eq:kc}
    \cos\kappa_c=\frac{V-2t_{\rm eff}}{4J_{\rm eff}}.
\end{equation}
The ME energy is $E_c=\varepsilon(\kappa_c)$. Substituting
Eq.~\eqref{eq:kc} into Eq.~\eqref{eq:disp} and simplifying
(Appendix~\ref{app:pm}) gives the central result of this work,
\begin{equation}\label{eq:ME}
    \;E_c=\frac{V^2-2V\,t_{\rm eff}}{4J_{\rm eff}}-2J_{\rm eff}\;.
\end{equation}
Equation~\eqref{eq:ME} describes a parabola in the energy--potential plane
with curvature $1/(4J_{\rm eff})$ and vertex at $V=t_{\rm eff}$,
$E_c^{\rm min}=-t_{\rm eff}^2/(4J_{\rm eff})-2J_{\rm eff}$. The vertex lies
below the bottom of the effective band, since the difference between the
two is $-(t_{\rm eff}-4J_{\rm eff})^2/(4J_{\rm eff})\le0$. Consequently, the
physical (rising) branch of the parabola, $V\ge t_{\rm eff}$, enters the
spectrum through the lower band edge: as $V$ increases, the ME sweeps
upward through the band, states below the ME localize, and states above it
remain extended. The matching momentum $\kappa_c$ is real for
$|V-2t_{\rm eff}|\le4J_{\rm eff}$, so the ME exists for
$V\in[2t_{\rm eff}-4J_{\rm eff},\;2t_{\rm eff}+4J_{\rm eff}]$;
outside this interval all states are either extended ($V$ too small) or
localized ($V$ too large), and the interval between these two bounds
defines the mixed phase in which extended and localized states coexist.
The two nonreciprocity parameters act geometrically on the
parabola: NN nonreciprocity shifts the vertex to larger $V$ through
$t_{\rm eff}$, whereas NNN nonreciprocity reduces the curvature through
$J_{\rm eff}$ and thereby widens the mixed phase. As a point-matching estimate,
Eq.~\eqref{eq:ME} is expected to be accurate when the NNN hopping is a
perturbation to the NN hopping, $J_{\rm eff}\ll t$, which is the regime
considered in the numerical tests below.

In the reciprocal limit $\beta=\delta=0$, Eq.~\eqref{eq:ME} becomes
$E_c=(V^2-2tV)/(4J)-2J$, the point-matching ME of the Hermitian generalized
AA model with NNN hopping. This expression is structurally different from
the result of Biddle \emph{et al.}~\cite{Biddle2009,Biddle2011},
\begin{equation}\label{eq:biddle}
    E_c^{\rm H}=\frac{V}{2}\!\left(\frac{t}{J}+\frac{J}{t}\right)
    -\frac{t^2}{J},
\end{equation}
which was obtained from a Lyapunov-exponent analysis in the limit $t\gg J$.
As shown below (Fig.~\ref{fig:herm}), the two expressions give numerically
similar boundaries for $J\ll t$. Both are analytical approximations to the
same fractal localization boundary, for which no exact closed form is
known; the point-matching form Eq.~\eqref{eq:ME} is the one that
generalizes to the nonreciprocal case.

%==============================================================================
\section{Numerical results}
\label{sec:num}

We test the analytical predictions by exact diagonalization of
Eq.~\eqref{eq:H} under PBC with $t=1$ and $J=0.1$, for which
$J_{\rm eff}\ll t$ and the point-matching estimate is expected to hold. All
system sizes are Fibonacci numbers, which provide the optimal rational
approximants to the golden-ratio frequency $\alpha$: the Hermitian baseline
is computed at $L=987$, the nonreciprocal phase diagrams at $L=610$, the
winding number at $L=377$, and the complex spectra at $L=987$ (PBC) and
$L=89$ (OBC). In all phase
diagrams the analytical ME is drawn along its physical branch where it lies
within the spectrum at the same potential strength.

\begin{figure}[tb]
    \centering
    \includegraphics[width=0.46\textwidth]{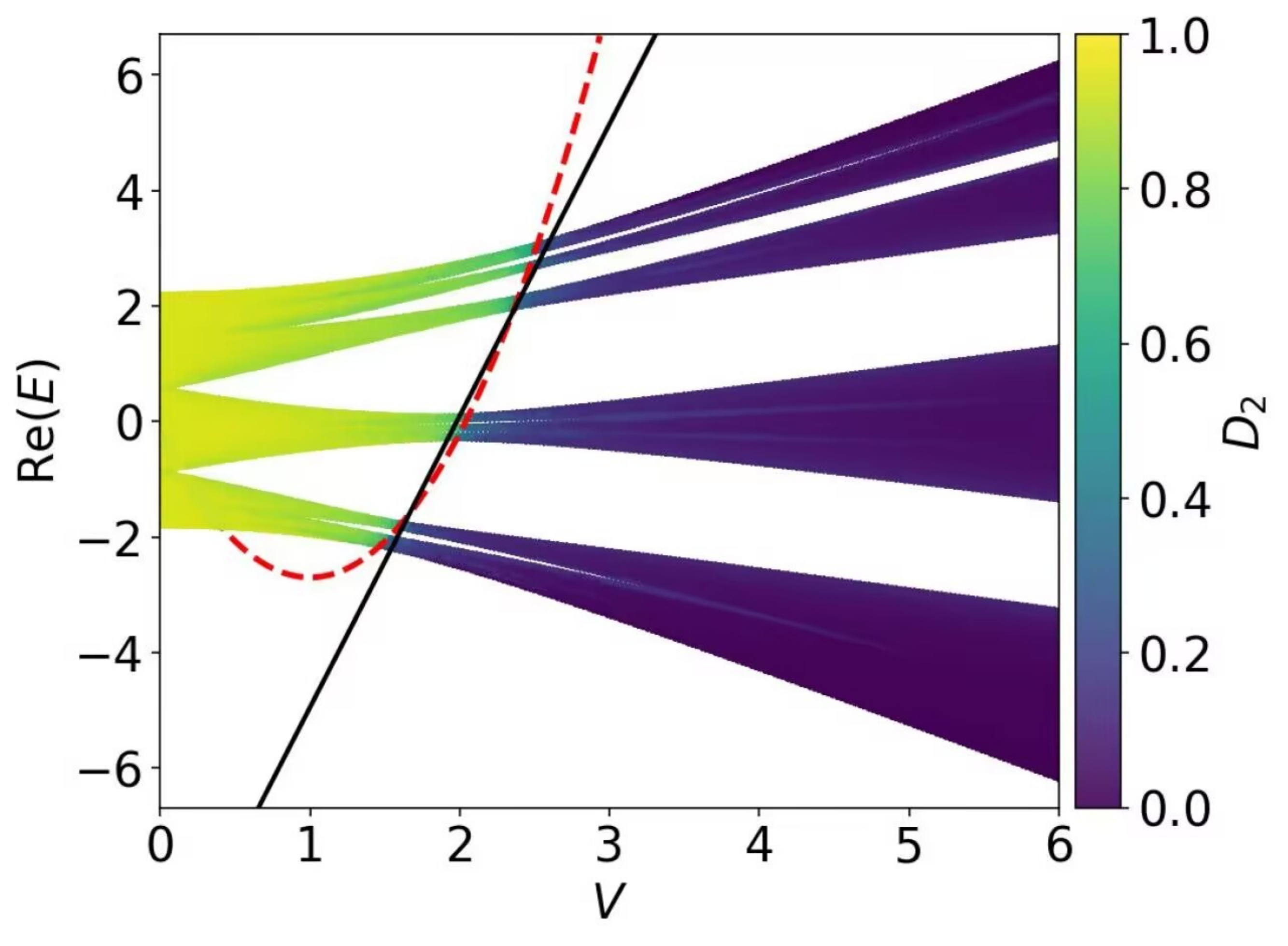}
    \caption{Hermitian limit $\beta=\delta=0$ ($t=1$, $J=0.1$, $L=987$,
    PBC). The color scale shows the fractal dimension $D_2$ (bright:
    extended, $D_2\to1$; dark: localized, $D_2\to0$). The Biddle ME of
    Eq.~\eqref{eq:biddle} (black solid line) and the point-matching ME of
    Eq.~\eqref{eq:ME} (red dashed line) are two analytical approximations to
    the same extended--localized boundary; both follow the numerical
    transition closely for $J\ll t$.}
    \label{fig:herm}
\end{figure}

Figure~\ref{fig:herm} shows the Hermitian phase diagram in the
energy--potential plane. For small $V$ all states are extended
($D_2\to1$); as $V$ increases, the spectrum localizes progressively, and
the energy-dependent boundary reflects the breaking of self-duality by the
NNN hopping. Both the Biddle line of Eq.~\eqref{eq:biddle} and the
point-matching parabola of Eq.~\eqref{eq:ME} track this boundary as it
rises through the spectrum from the lower band edge. The two curves are
numerically close but not identical, in accordance with their different
analytical origins, and they agree best in the central part of the
spectrum, where the $t\gg J$ approximation is most reliable. This figure
establishes the Hermitian baseline against which the nonreciprocal shifts
are measured.

\begin{figure*}[tb]
    \centering
    \includegraphics[width=0.96\textwidth]{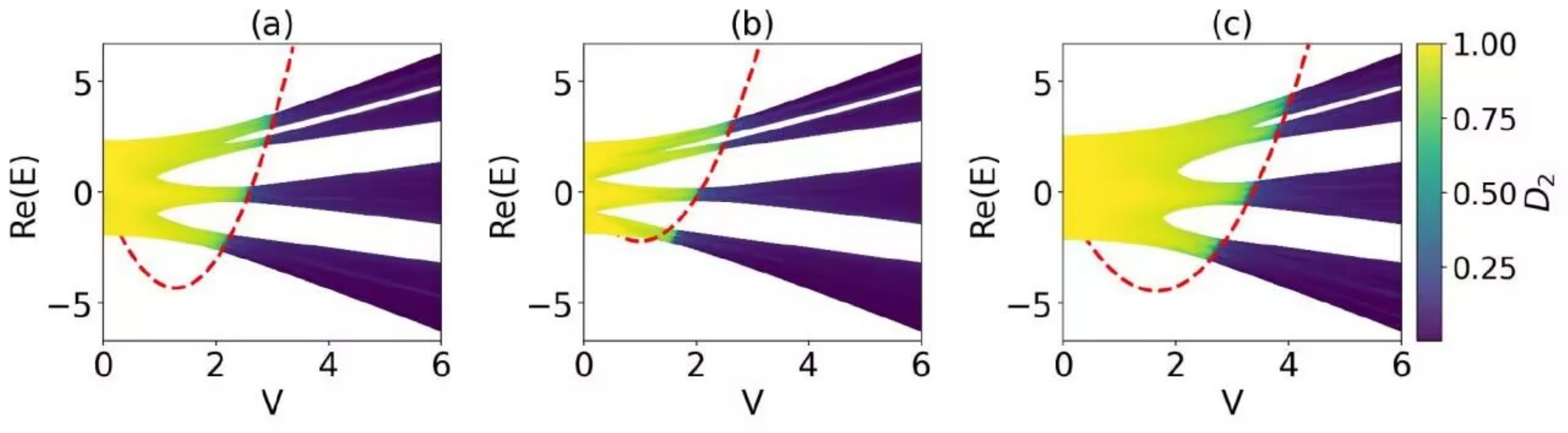}
    \caption{Nonreciprocal phase diagrams ($t=1$, $J=0.1$, $L=610$, PBC)
    together with the point-matching ME of Eq.~\eqref{eq:ME} (red dashed
    lines). (a) NN nonreciprocity only, $\beta=0.25$, $\delta=0$
    ($t_{\rm eff}=e^{0.25}\approx1.28$, $J_{\rm eff}=0.1$): the boundary
    shifts rigidly to larger $V$. (b) NNN nonreciprocity only, $\beta=0$,
    $\delta=0.25$ ($t_{\rm eff}=1$,
    $J_{\rm eff}=0.1\,e^{0.25}\approx0.13$): the curvature of the boundary
    decreases and the parabola widens. (c)
    Combined case,
    $\beta=\delta=0.5$: the shift and the widening superpose. In all three
    panels the analytical parabola, with no adjustable parameters, follows
    the extended--localized boundary.}
    \label{fig:cases}
\end{figure*}

Turning to the nonreciprocal regime, Fig.~\ref{fig:cases} presents the
three physically distinct configurations. In panel (a), only the NN hopping is
nonreciprocal ($\beta=0.25$, $\delta=0$), so that
$t_{\rm eff}=t\,e^{0.25}\approx1.28$ while $J_{\rm eff}=J$. The boundary
shifts rigidly toward larger potential strength: the vertex of the parabola
moves from $V=1$ (the Hermitian value) to $V=t_{\rm eff}\approx1.28$, while
the curvature $1/(4J_{\rm eff})=2.5$ is unchanged. Physically, the
nonreciprocal NN hopping enhances the effective bandwidth, so a stronger
potential is required to localize the states.

In panel (b), only the NNN hopping is nonreciprocal ($\beta=0$,
$\delta=0.25$), so that $t_{\rm eff}=t$ and
$J_{\rm eff}=J\,e^{0.25}\approx0.13$. The vertex position is unchanged at
$V=1$, but the curvature decreases from $2.5$ to
$1/(4\times0.13)\approx1.9$, which widens the parabola and broadens the
range of potential strengths over which extended and localized states
coexist. The same parameter set is used for the spectral-topology analysis
of Fig.~\ref{fig:nhse} below.

Panel (c) combines both nonreciprocities ($\beta=\delta=0.5$), and the
resulting boundary is simultaneously shifted and widened. In all three
configurations, the single formula Eq.~\eqref{eq:ME}, with no adjustable
parameters, reproduces the numerically determined extended--localized
boundary across the spectrum. A quantitative comparison at fixed energies
shows that the predicted critical potential agrees with the numerical
boundary to within a few percent over the central and upper parts of the
spectrum, while the largest deviations, of order ten percent, occur near
the lower band edge. The same trend is already present in the Hermitian
limit (Fig.~\ref{fig:herm}), reflecting the fact that the point-matching
construction anchors the boundary at the high-symmetry points of the band
and is least constrained at its edges.

\begin{figure*}[tb]
    \centering
    \includegraphics[width=0.94\textwidth]{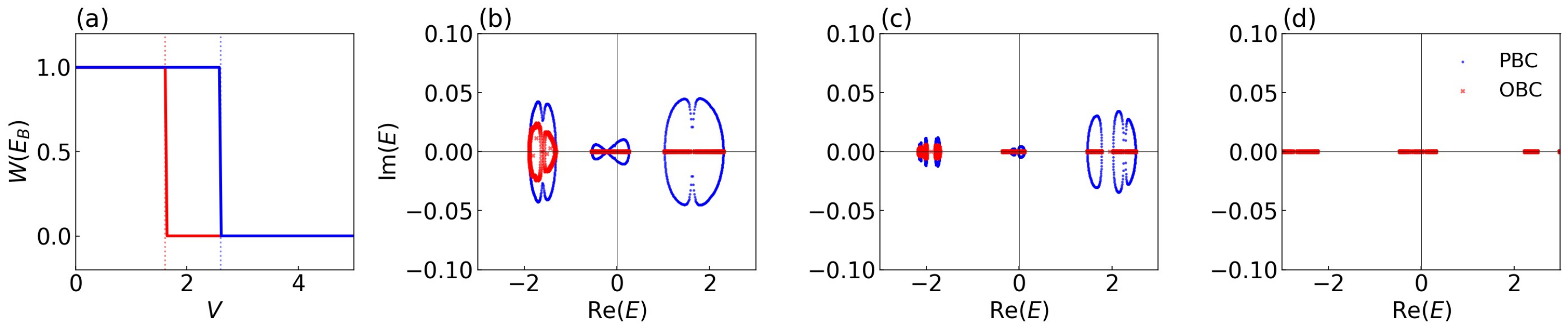}
    \caption{Skin-effect topology for NNN nonreciprocity ($\beta=0$,
    $\delta=0.25$, $t=1$, $J=0.1$). (a) Spectral winding numbers
    $W(E_B)$ at two base energies near the band edges,
    $E_B=-2.0$ (red) and $E_B=2.8$ (blue), as a function of $V$.
    The winding number at the lower band edge drops at
    $V_1\approx1.6$, marking the onset of the mixed phase where the
    ME enters the spectrum and the first localized states appear. The
    winding number at the upper band edge drops at $V_2\approx2.6$,
    marking the end of the mixed phase where the last extended states
    localize. The interval $[V_1,V_2]$ defines the mixed phase in which
    extended and localized states coexist at different energies.
    (b)--(d) Complex energy spectra under PBC (blue dots) and OBC
    (red crosses) at $V=0.8$, $1.5$, and $3.0$ ($L=987$).
    At $V=0.8$ the PBC spectrum forms loops that enclose the origin
    while the OBC spectrum collapses onto the real axis, which is the
    spectral hallmark of the NHSE. At $V=1.5$, near the onset of the
    mixed phase, the loops are shrinking.
    At $V=3.0$ nearly the entire spectrum is real under both boundary
    conditions; the NHSE is suppressed and all states are localized.}
    \label{fig:nhse}
\end{figure*}

Figure~\ref{fig:nhse} examines the fate of the NHSE as the potential
localizes the spectrum, for the same parameters as in
Fig.~\ref{fig:cases}(b) ($\beta=0$, $\delta=0.25$). Panel (a) shows
the spectral winding numbers $W(E_B)$ evaluated at two base energies
near the band edges, $E_B=-2.0$ and $E_B=2.8$, as a function of $V$.
For small $V$ both winding numbers equal one: the complex PBC spectrum
forms loops that enclose both base energies and the NHSE is active at
those energies. As $V$ increases, the energy-dependent ME enters the
spectrum from the lower band edge, and the winding number at the lower
base energy $E_B=-2.0$ drops to zero at $V_1\approx1.6$, marking the
onset of the mixed phase. The winding number at the upper base energy
$E_B=2.8$ drops at $V_2\approx2.6$, marking the end of the mixed phase
where the last extended states at the upper band edge localize.
The interval $[V_1,V_2]$ thus defines the mixed phase in which
localized and extended states coexist at different energies, a
characteristic feature of systems with energy-dependent MEs.

Panels (b)--(d) display the corresponding complex spectra. At $V=0.8$
[panel (b)], the PBC spectrum forms loops that enclose the origin, while
the OBC spectrum collapses onto the real axis; this extreme sensitivity of
the spectrum to the boundary conditions is the defining signature of the
NHSE. At $V=1.5$ [panel (c)], near the onset of the mixed phase, the
spectral loops are shrinking. At $V=3.0$ [panel (d)], well above $V_2$,
nearly the entire spectrum is real under both boundary conditions; the
NHSE is suppressed and all states are localized.

The physical picture is as follows. Extended states sample the entire
ring and experience the net nonreciprocal flow, thereby acquiring complex
energies; localized states are confined to a region much smaller than the
ring and are insensitive to the boundary, so their energies remain real.
The vanishing of the eigenvalue imaginary parts at a given energy
therefore coincides with the localization of the corresponding states and
is predicted quantitatively by the ME of Eq.~\eqref{eq:ME}. The two
winding numbers at the band-edge base energies directly bracket the
mixed phase: $V_1$ marks the appearance of the first localized states at
the lower band edge, while $V_2$ marks the localization of the last
extended states at the upper band edge.

%==============================================================================
\section{Conclusion}
\label{sec:concl}

We have derived a closed-form mobility edge for the non-Hermitian
generalized Aubry-Andr\'{e} model with nonreciprocal nearest-neighbor and
next-nearest-neighbor hopping by extending the Fermi-surface
point-matching method to nonreciprocal systems. The central result is an
effective-hopping description: the ME is the Hermitian point-matching
parabola with the bare hoppings replaced by
$t_{\rm eff}=t\,e^{|\beta|}$ and $J_{\rm eff}=J\,e^{|\delta|}$. The skin
effect amplifies the apparent hopping, so a stronger potential is required
to localize the states, and the ME shifts to larger $V$. Because only
$|\beta|$ and $|\delta|$ enter, the direction of the nonreciprocity does
not affect the bulk ME. The effective hoppings also account for the
solvable limits: $J_{\rm eff}\to0$ gives the nonreciprocal AA chain
($V_c=2t_{\rm eff}$)~\cite{Hatano1996,Longhi2019}, while
$t_{\rm eff}\to0$ gives decoupled nonreciprocal sublattices
($V_c=2J_{\rm eff}$); on the line $\delta=2\beta$ an imaginary gauge
transformation removes the nonreciprocity under OBC, but the
transformation is not single valued on a ring, so the PBC bulk ME
remains shifted (Appendix~\ref{app:gauge}). Two caveats delimit the
validity: the generalized AA model is not exactly self-dual and its
spectrum is fractal, so Eq.~\eqref{eq:ME} is an analytical estimate
accurate for $J_{\rm eff}\ll t$; and the independent treatment of
$|\beta|$ and $|\delta|$ applies to the bulk ME under PBC, whereas the
OBC properties depend on $\tilde\delta=\delta-2\beta$.

Exact diagonalization confirms the analytical boundary for all three
nonreciprocal configurations and recovers the Hermitian result in the
reciprocal limit. The analysis of the spectral topology shows that the spectral winding
numbers evaluated at base energies near the two band edges directly
bracket the mixed phase, dropping at the potential strengths where the
energy-dependent ME enters and exits the spectrum.

Non-Hermitian quasiperiodic lattices have been realized in optical fiber
loops~\cite{Weidemann2020,WeidemannNature2022}, photonic quantum
walks~\cite{Xiao2020,Lin2022}, topolectrical circuits~\cite{Helbig2020},
active mechanical metamaterials~\cite{Brandenbourger2019,Ghatak2020}, and
ultracold atomic gases~\cite{Liang2022}. The ME predicted here can be
probed through the energy-resolved participation ratio, and the winding
number through the boundary dependence of the spectrum. Natural extensions
include longer-range nonreciprocal hopping, higher dimensions, and the
interplay of energy-dependent MEs with Floquet
driving~\cite{Zhou2021,WeidemannNature2022} and many-body
interactions~\cite{Hamazaki2019,Qian2024}.

\begin{acknowledgments}
This work was supported by the Natural Science Foundation of Jiangsu
Province (Grant No.~BK20231320).
\end{acknowledgments}

\section*{DATA AVAILABILITY}
The data that support the findings of this article are not publicly available. The data are available from the authors upon reasonable request.

%==============================================================================
\appendix

\section{Derivation of the point-matching mobility edge}
\label{app:pm}

For $V=0$ and PBC, the Bloch ansatz $\psi_i\propto e^{i\kappa i}$ applied
to Eq.~\eqref{eq:H} gives
$E(\kappa)=2t\cos(\kappa-i\beta)+2J\cos(2\kappa-i\delta)$. The imaginary
gauge transformation $c_i\to e^{-\beta i}c_i$ removes the NN
nonreciprocity and shifts the NNN one to $\tilde\delta=\delta-2\beta$
(Appendix~\ref{app:gauge}); the amplification of the NN and NNN hopping
experienced by a state propagating around the ring has magnitude
$e^{|\beta|}$ and $e^{|\delta|}$, respectively. Absorbing these factors
into the amplitudes yields the effective dispersion of
Eqs.~\eqref{eq:disp} and \eqref{eq:teff}.

Writing $x=\cos\kappa$ and using $\cos2\kappa=2x^2-1$, the effective
dispersion reads $\varepsilon=4J_{\rm eff}x^2+2t_{\rm eff}x-2J_{\rm eff}$.
The point-matching ansatz~\cite{Vu2023} identifies the dual hopping $V$
with the effective NN hopping renormalized by the NNN term at the critical
momentum, $V=2t_{\rm eff}+4J_{\rm eff}x_c$, which gives
$x_c=(V-2t_{\rm eff})/(4J_{\rm eff})$, i.e., Eq.~\eqref{eq:kc}.
Substituting into $E_c=\varepsilon(\kappa_c)$,
\begin{align}
    E_c&=4J_{\rm eff}x_c^2+2t_{\rm eff}x_c-2J_{\rm eff}\nonumber\\
    &=\frac{(V-2t_{\rm eff})^2}{4J_{\rm eff}}
     +\frac{t_{\rm eff}(V-2t_{\rm eff})}{2J_{\rm eff}}-2J_{\rm eff}
     \nonumber\\
    &=\frac{V^2-2Vt_{\rm eff}}{4J_{\rm eff}}-2J_{\rm eff},
\end{align}
which reproduces Eq.~\eqref{eq:ME}. The matching momentum $\kappa_c$ is
real for $|x_c|\le1$, i.e., for $|V-2t_{\rm eff}|\le4J_{\rm eff}$; outside
this window $\kappa_c$ becomes complex and the ME continues analytically.
The physical branch is the rising branch with $V\ge t_{\rm eff}$.

\section{Solvable limits $J=0$ and $t=0$}
\label{app:limits}

\begin{figure}[tb]
    \centering
    \includegraphics[width=0.48\textwidth]{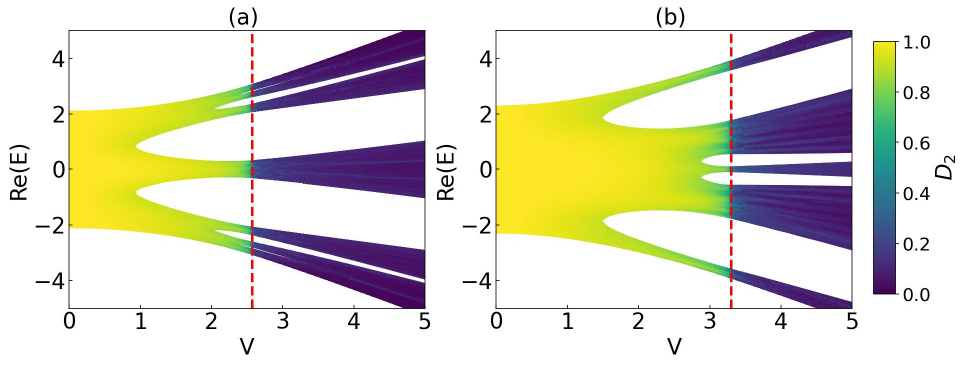}
    \caption{Solvable limits ($L=610$, PBC). (a) $J=0$, $t=1$,
    $\beta=0.25$: the localization transition is energy independent and
    occurs at $V_c=2t\,e^{|\beta|}\approx2.57$ (red dashed vertical line).
    (b) $t=0$, $J=1$, $\delta=0.5$: the transition is energy independent
    and occurs at $V_c=2J\,e^{|\delta|}\approx3.30$ (red dashed vertical
    line). In both limits the ME degenerates into a vertical line, and no
    energy-dependent boundary exists. In (b) each eigenstate extends over a
    single sublattice, so the fractal dimension of extended states is
    slightly reduced, $D_2\approx1-\ln2/\ln L$.}
    \label{fig:limits}
\end{figure}

\paragraph{Nonreciprocal AA chain ($J=0$).}
The eigenvalue equation reduces to
$E\psi_i=V_i\psi_i+t\,e^{\beta}\psi_{i-1}+t\,e^{-\beta}\psi_{i+1}$, with
$V_i=V\cos(2\pi\alpha i+\phi)$. The imaginary gauge transformation
$\psi_i=e^{-\beta i}u_i$ maps this equation to the Hermitian AA model
$Eu_i=V_iu_i+t(u_{i-1}+u_{i+1})$, whose Lyapunov exponent
$\gamma=\max[0,\ln(V/2t)]$ is independent of energy. On a ring, an
eigenstate of the Hermitian model survives the gauge transformation only
if $\gamma>|\beta|$, so the global transition shifts to
$V_c=2t\,e^{|\beta|}$ [Fig.~\ref{fig:limits}(a)], an energy-independent
vertical line without an ME, in agreement with
Refs.~\cite{Hatano1996,Longhi2019,Jiang2019}.

\paragraph{Decoupled sublattices ($t=0$).}
Setting $t=0$ in Eq.~\eqref{eq:H} decouples the even and odd sublattices.
For the even sublattice, with $n=2m$,
\begin{equation}
    Eu_{2m}=V\cos(4\pi\alpha m+\phi)\,u_{2m}
        +J\,e^{\delta}u_{2m-2}+J\,e^{-\delta}u_{2m+2},
\end{equation}
and analogously for the odd sublattice with a shifted phase. Each
sublattice is a nonreciprocal AA chain with hopping $J$, nonreciprocity
$\delta$, and frequency $2\alpha$. The gauge argument of the $J=0$ case,
with $t\to J$ and $\beta\to\delta$, gives $V_c=2J\,e^{|\delta|}$
[Fig.~\ref{fig:limits}(b)]. Both limits are degenerations of
Eq.~\eqref{eq:ME}: as $J_{\rm eff}\to0$ or $t_{\rm eff}\to0$ the parabola
collapses to a vertical line at $V=2t_{\rm eff}$ or $V=2J_{\rm eff}$,
respectively.

\section{Imaginary gauge transformation and the line $\delta=2\beta$}
\label{app:gauge}

\begin{figure}[tb]
    \centering
    \includegraphics[width=0.46\textwidth]{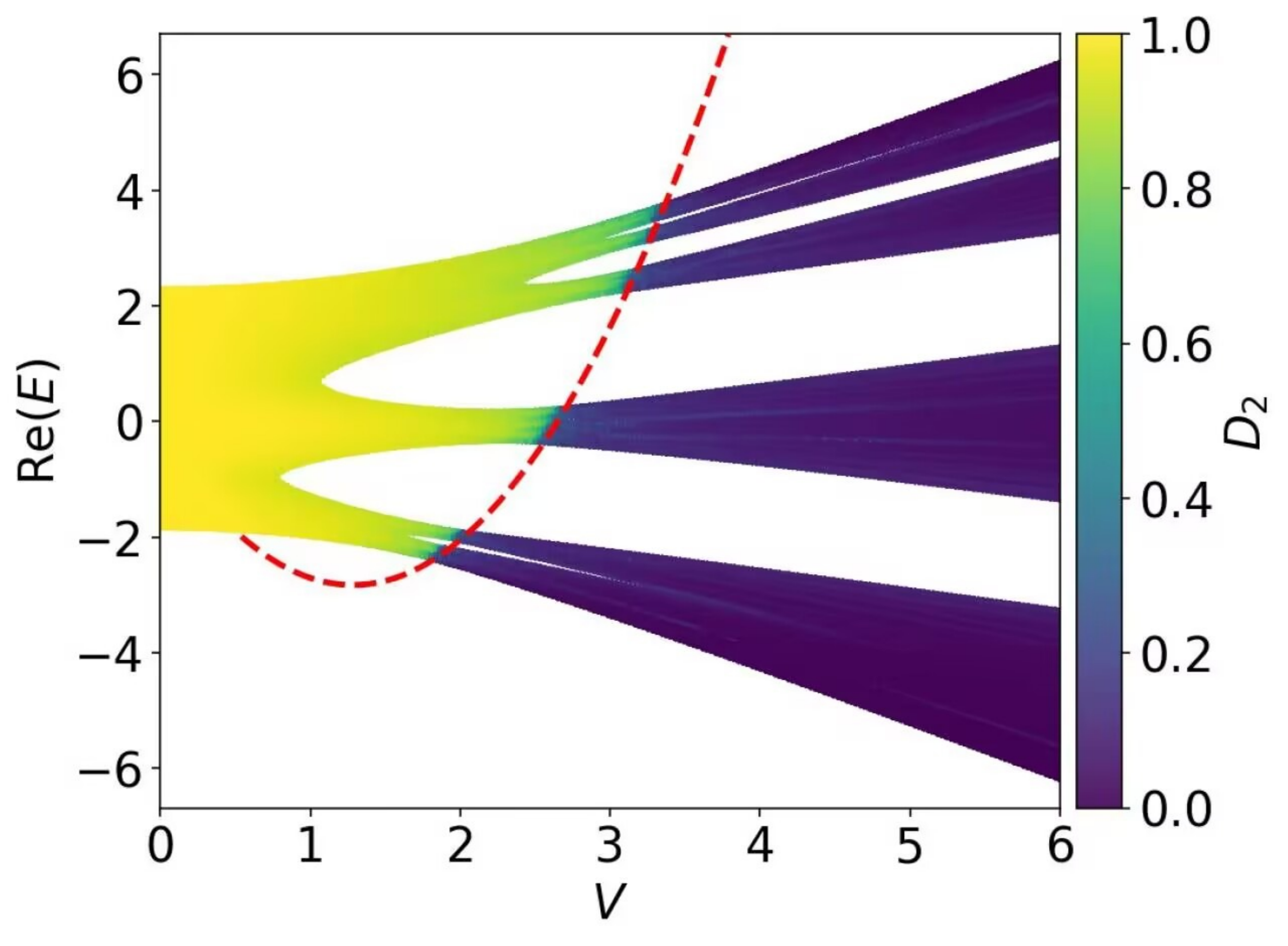}
    \caption{Gauge-related case $\delta=2\beta$ ($\beta=0.25$,
    $\delta=0.50$, $\tilde\delta=0$; $t=1$, $J=0.1$, $L=610$, PBC). The
    color scale shows the fractal dimension $D_2$. The red dashed line is
    the point-matching ME of Eq.~\eqref{eq:ME} with
    $t_{\rm eff}=e^{0.25}$ and $J_{\rm eff}=0.1\,e^{0.50}$. Although the
    imaginary gauge transformation maps the open chain to the Hermitian
    model, it is not single valued on the PBC ring: the bulk boundary is
    shifted to larger $V$ relative to the Hermitian Biddle line [whose
    $E=0$ crossing lies at $V\approx1.98$, versus $V\approx2.65$ for the
    parabola] and is well described by the effective-hopping parabola.}
    \label{fig:gauge}
\end{figure}

Inserting $\psi_i=e^{-\beta i}u_i$ into the eigenvalue equation of
Eq.~\eqref{eq:H} yields
\begin{equation}
    Eu_i=V_iu_i+t(u_{i+1}+u_{i-1})
        +J\,e^{\tilde\delta}u_{i+2}+J\,e^{-\tilde\delta}u_{i-2},
\end{equation}
with the residual NNN nonreciprocity $\tilde\delta=\delta-2\beta$: the NN
hopping becomes reciprocal, the potential is unchanged, and only the NNN
hopping retains nonreciprocity. On the line $\delta=2\beta$ one has
$\tilde\delta=0$, and the transformed model is Hermitian.

Under OBC the transformation is an exact similarity transformation, so the
OBC localization properties on this line are those of the Hermitian model.
On a ring, however, the factor $e^{-\beta i}$ is not single valued: the
gauge maps PBC onto a chain with a boundary defect of strength
$e^{\mp\beta L}$, and the Hatano-Nelson
argument~\cite{Hatano1996,Hatano1998,Longhi2019} shows that an eigenstate
of the Hermitian model with Lyapunov exponent $\gamma_{\rm H}(E)$ survives
on the ring only if $\gamma_{\rm H}(E)>|\beta|$; states with
$\gamma_{\rm H}(E)<|\beta|$ delocalize around the ring and acquire complex
energies. The PBC mobility edge on the line $\delta=2\beta$ is therefore
determined by
\begin{equation}\label{eq:gHbeta}
    \gamma_{\rm H}(E_c)=|\beta|,
\end{equation}
rather than by $\gamma_{\rm H}=0$. Since $\gamma_{\rm H}(E)$ vanishes
linearly at the Hermitian ME,
$\gamma_{\rm H}(E)\simeq\kappa(E-E_c^{\rm H})$ with
$\kappa=d\gamma_{\rm H}/dE|_{E_c^{\rm H}}>0$, the leading nonreciprocal
correction is
\begin{equation}\label{eq:pert}
    E_c\approx E_c^{\rm H}+\frac{|\beta|}{\kappa}.
\end{equation}
Because $\gamma_{\rm H}(E)$ of the NNN model has no simple closed form,
$\kappa$ must be evaluated numerically, and the effective-hopping parabola
of Eq.~\eqref{eq:ME} provides the practical closed-form alternative.
Figure~\ref{fig:gauge} confirms this picture for $\beta=0.25$,
$\delta=0.50$: the PBC fractal-dimension boundary is shifted away from the
Hermitian Biddle line (whose $E=0$ crossing lies at $V\approx1.98$) and
follows Eq.~\eqref{eq:ME}, whose crossing of $E=0$ at $V\approx2.65$
agrees with the numerical boundary of $D_2$.

\section{Numerical methods}
\label{app:num}

The phase diagrams are obtained by exact diagonalization of
Eq.~\eqref{eq:H} under PBC for Fibonacci sizes $L$, which provide the best
rational approximants to the frequency $\alpha$. The fractal dimension
$D_2$ is computed from the normalized right eigenvectors via
Eq.~\eqref{eq:D2}. For Hermitian parameters the spectrum is obtained with
a Hermitian eigensolver; for non-Hermitian parameters a full complex
eigendecomposition is performed. PBC is used throughout so that $D_2$
reflects bulk localization rather than the skin effect, and the numerical
ME is identified as the boundary where $D_2$ drops from values near one to
values near zero. The phase diagrams in
Figs.~\ref{fig:herm}, \ref{fig:cases}, \ref{fig:limits}, and
\ref{fig:gauge} are computed at a fixed representative phase $\phi=0$.

The winding number of Eq.~\eqref{eq:wind} is computed by discretizing
$\theta\in[0,2\pi]$ into $N\ge120$ points, distributing the flux uniformly
over the ring (each NN link acquires a factor $e^{i\theta/L}$ and each NNN
link a factor $e^{2i\theta/L}$), evaluating
$\det[H(\theta_j)-E_B]$ through a stabilized LU factorization, unwrapping
the accumulated argument, and dividing the total phase winding by $2\pi$.
The result is integer valued for converged $N$ and is robust with respect
to the system size and the potential phase. In Fig.~\ref{fig:nhse}(a) the
winding numbers are evaluated at two base energies near the band edges,
$E_B=-2.0$ and $E_B=2.8$; their drop points $V_1$ and $V_2$ are located
by bisection in $V$ to an accuracy of $10^{-3}$. The complex spectra in
Fig.~\ref{fig:nhse} are obtained by direct diagonalization at $L=987$
under PBC and $L=89$ under OBC; the small OBC size is dictated by the
condition number, of order $e^{|\tilde\delta|L/2}$, of the similarity
transformation connecting the open chain to its reciprocal counterpart,
which renders the OBC spectrum of a long nonreciprocal chain numerically
unreliable in double precision.

\nocite{apsrev42Control}
\bibliographystyle{apsrev4-2}
\bibliography{references}

\end{document}